\documentclass[aps,apl,showpacs,amsfonts,amssymb,manuscript]{revtex4}

\usepackage[dvips]{graphicx}

\begin{document}

\title{Electric field effect on BiFeO$_3$ single crystal investigated by Raman spectroscopy}

\author{M. Cazayous}
\email{maximilien.cazayous@espci.fr}
\author{D. Malka}
\affiliation{Laboratoire Mat\'eriaux et Ph\'enom\`enes Quantiques (UMR 7162), 
Universit\'e Paris Diderot-Paris 7, 75205 Paris cedex 13, France\\
Laboratoire de Physique des Solides (UPR5 CNRS),
Ecole Superieure de Physique et de Chimie Industrielle, 10 rue Vauquelin, 
75231 Paris, France}

\author{D. Lebeugle}
\author{D. Colson}
\affiliation{Service de Physique de l'Etat Condens\'e, DSM/DRECAM/SPEC, CEA Saclay, 91191 Gif-sur-Yvette, France}

\date{\today}

\begin{abstract}
Micro-Raman spectroscopy has been used to study lattice dynamics associated with the ferroelectric domains of a BiFeO$_3$ single crystal at low temperature. The phonon assignment shows a large frequency splitting between the transverse and longitudinal components of the A$_1$ phonon mode related to the Bi-O bonds in contrast with thin films where the splitting is negligible. 
Applying an external electric field induces frequency shifts of the low energy 
modes related to the Bi-O bonds. These softenings are due to a tensile stress via the piezoelectric effect. 
We give estimates of the phonon deformation potentials.
\end{abstract}


\maketitle

Multiferroics have been of particular interest for understanding the fundamental aspects of the  
mechanism that gives rise to magneto-ferroelectric coupling, as well as for developping
devices including spintronics, electro-optics and data 
storage.\cite{Eerenstein} BiFeO$_3$ (BFO) is an attractive and intriguing material because it exhibits 
both ferroelectric and antiferromagnetic orders at room temperature. 
First observed in thin films\cite{Wang}, a strong ferroelectric polarization is also now measured in single crystal.\cite{Lebeugle,Lebeugle1}

The polarized domain structures in single crystal ferroelectric materials 
could be an obstacle for devices. Indeed, the dielectric 
and electromechanical properties are very sensitive to the ferroelectric domains which depend on 
an external electric field or mechanical stress.\cite{Bonda} The electric field effects on the ferroelectric properties of BiFeO$_3$ is a topic of current interest.\cite{Zhao}
\par
Despite an intense study of BFO thin films, Raman measurements remain scarce \cite{Singh1,Singh2} especially on single crystal \cite{Haumont}. 
The first step needed to understand the ferroelectricity is to investigate the lattice dynamic. It
plays a key role in the coupling between the ferroelectric and magnetic orders and its behavior under stress or an applied electric field\cite{Bark}.  
\par
In this letter, we report polarized micro-Raman measurements on ferroelectric domains 
in BiFeO$_3$ single crystal at 4~K. We have assigned the vibrationnal modes and investigated in a reversible way 
the structural changes induced by the application of an external electric field.
\par
BiFeO$_3$ single crystals were grown in air using a Bi$_2$O$_3$-Fe$_2$O$_3$ flux technique.\cite{Lebeugle} 
The as-grown crystals are millimetre sized black platelets. Electron microprobe analysis confirms the stoichiometry of BiFeO$_3$. The mean thickness of the analysed crystals is 10$\mu$m. 
BiFeO$_3$ single crystal is a ferroelectric with a Curie temperature, $T_c$$\sim$1100 K,
and shows antiferromagnetic order below the N\'eel temperature, $T_N$$\sim$640 K.\cite{Smolenski,Ismilzade} In its ferroelectric phase, BFO has a rhombohedrally distorted perovskite structure with the space group R3c and lattice constants a$_{hex}$=5.571\AA~and c$_{hex}$=13.868\AA~at room temperature.\cite{Kubel} The spontaneous polarization is along one of the  pseudo-cubic [111]  directions with a value between 50 and 100 $\mu$C/cm$^2$.\cite{Lebeugle}. 
The orientation of the crystal axes has been determined by X-Ray analysis.\cite{Lebeugle, Lebeugle1} 
BFO presents a G-type antiferromagnetic structure with a long range modulation associated with a cycloidal spiral.\cite{Sosnowska}
\par
Micro and macro Raman measurements have been performed 
with a laser spot size of 1 $\mu$m and 100 $\mu$m, respectively \cite{Sac}. We have used the 514.5 nm (2.41 eV) excitation 
line from a Ar$^+$-Kr$^+$ mixed gas laser. The Raman scattering spectra were recorded at 
10 and 300~K using a triple spectrometer Jobin Yvon T64000. 
The spectrometer was in substractive configuration with a resolution of 0.1 cm$^{-1}$ 
in order to detect the electrical field effects on phonon modes.
The electrical field has been applied along the z-axis of the crystal using top and bottom ohmic contacts formed by depositing indium drops to the sample, and annealing the sample at 240$^o$C for 10 min.
\par

\begin{figure}
\includegraphics*[width=18cm]{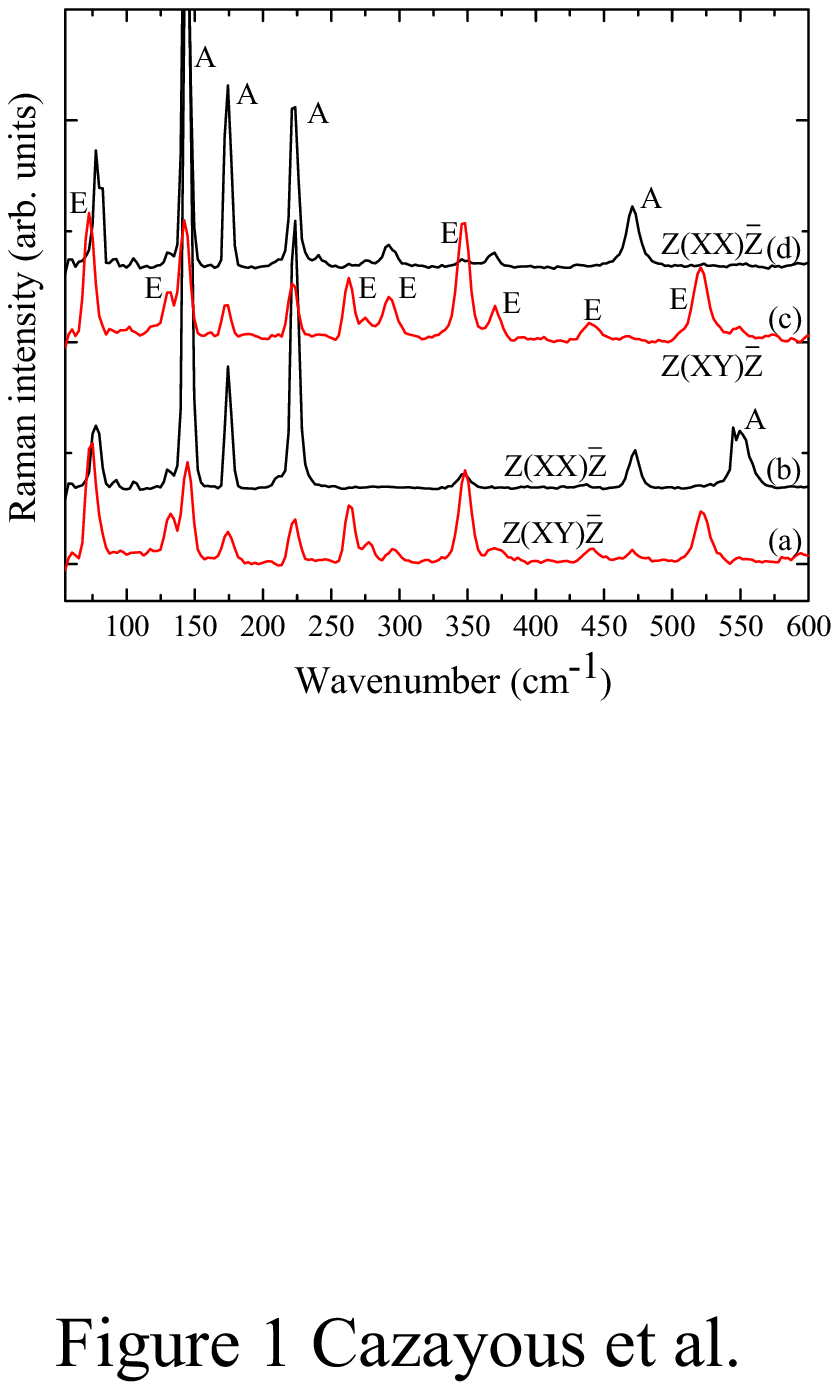}\\
\caption{\label{Figure1} Polarized Raman spectra (a), (b) and (c), (d) on two different single domains, respectively; 
(a) and (c) in $Z(XY)\bar{Z}$, (b)  and (d) in  $Z(XX)\bar{Z}$ backscattering geometry.}
\end{figure}

Figure \ref{Figure1} shows polarized spectra at 10~K obtained from the two different ferroelectric 
domains in $Z(XX)\bar{Z}$ and $Z(XY)\bar{Z}$ micro Raman geometries.\cite{Porto}
The (a),(b) spectra and (c), (d) ones are related to two different ferroelectric domains. 
Domains have been visually selected by polarized light. The contrast between bright domains 
(which correspond to a direction of the spontaneous polarization) and dark domains 
(which correspond to another one) comes from the birefringence property of the crystal.
\par
Let us focus first on the lattice dynamics. The vibrational modes detected by Raman spectrosocpy depend on i)  
the crystal symmetry which controls the matrix elements of  the Raman tensor and ii)  
the incident and scattered light polarizations which stress the Raman tensor.
Group theory predicts 13 Raman active phonons: $\Gamma=4A_1+9E$. 
All the measurements have been performed in backscattering configuration 
(incident wave vector parallel to the scattered one).  In order to get the pure E modes we have used 
crossed polarization along the x and y directions, this corresponds to $Z(XY)\bar{Z}$ geometry. 
The $A_1+E$ modes have been obtained using parallel polarizations along the x direction. This corresponds 
to the $Z(XX)\bar{Z}$ geometry.
In this scattering geometry, the propagation vector of the relevant phonon is along the Z direction which is 
parallel to the principal [111] polar axis. The eigenvectors of the A$_1$ normal mode are parallel to the Z 
direction whereas those of the E mode are along X and Y directions. The relevant A$_1$-symmetry phonons are therefore 
longitudinal optic (LO) while the E-symmetry phonons are transverse optic (TO).
A clear assignement of the Raman modes to a specific bond motion is still missing. An attempt 
in thin films has concluded that Bi-O bonds contribute mostly to A$_1$ modes, first 
and second-order E(TO) modes and Fe-O bonds to third and fourth-order E(TO) modes.\cite{Singh1} 
The changes from $Z(XY)\bar{Z}$ 
 to $Z(XX)\bar{Z}$ configurations should lead to an enhancement of the A$_1$ mode intensities 
in an opposite way of the E modes. 
There is no ambiguity about the four A$_1$ modes (145, 175.5, 222.7 and 471 cm$^{-1}$)
and eight out of nine E modes (132, 263, 276, 295, 348, 370, 441, 523 cm$^{-1}$).\cite{Singh1,Singh2,Massey} 
The peak at 550 cm$^{-1}$ is only observed with a strong signal 
in one ferroelectric domain and in parallel configuration [Fig. \ref{Figure1}(b)]. 
We assigned this peak to the TO forbidden vibrational mode of the 
the fourth A$_1$ mode. This assumption is supported by the observation 
of a peak centered around 1100 cm$^{-1}$ associated with the 2TO line.

We note that a negligible LO-TO splitting of A$_1$ modes has been observed in thin films \cite{Singh2} and has been interpreted as the domination of the short-range interatomic force over the long-range ionic force. 
However a giant LO-TO splitting in perovskite ferroelectrics has been measured and discussed in Ref.~\cite{Zhong} and especially for BFO in Ref.~\cite{Hermet}. 
The observe splitting comes from a strong Coulomb interaction and a high sensitivity of the ferroelectricity to the domain structure. A small LO-TO splitting thus appears to be a thin film property, in contrast with the large one observed in our measurement on single crystals. 
A clear understanding of the LO-TO splitting and phonon assigment need more theoretical contributions 
via ab-initio calculations.\cite{Hermet}.
\par
In our case, we observe only the A$_1$(TO) mode in 
a ferroelectric domain [peak at 550 cm$^{-1}$ in (b) spectrum of Fig. \ref{Figure1}] which is three 
times larger than the spot diameter. One possible origin of the observation of the A$_1$(TO) 
mode is the local symmetry breaking due to strain fields of the domain 
walls which extend inside the domain making the Raman-forbidden phonons active. 
\par
The peaks at 75 and 81 cm$^{-1}$ [in spectra (a), (c) and (b), (d) of Fig. 1, respectively] 
have been assigned to the last missing E mode. In particular, no enhancement of their 
intensity is observed as one goes from the parallel to the cross polarizations in contrast to the A$_1$ modes. 
The mode at 81 cm$^{-1}$ is detected in $Z(XX)\bar{Z}$ configuration and is assigned to the first-order E(LO) mode. 
The mode at 75 cm$^{-1}$ corresponds to the first-order E(TO) mode.
Our assignment is in good agreement with the two modes at the same fequencies observed by infrared measurements on BiFeO3 ceramics.\cite{Kamba}
Our samples are pure and no phonon modes associated with impurity phases 
(Bi$_2$O$_3$, Fe$_2$O$_3$, Fe$_3$O$_4$, ...) have been detected.
The Raman peak assignment is summarized in Table \ref{assignment}. 
Singh et al.~\onlinecite{Singh2} have reported on BFO films, similar vibrational modes around identical frequencies: E(TO) 
modes at 275, 335, 363, 456 and 549 cm$^{-1}$ and  A$_1$(LO) modes at 136, 168 and 212 cm$^{-1}$. 
In the present work, we have observed all the modes predicted by the group theory.

\par
\begin{figure}
\includegraphics*[width=18cm]{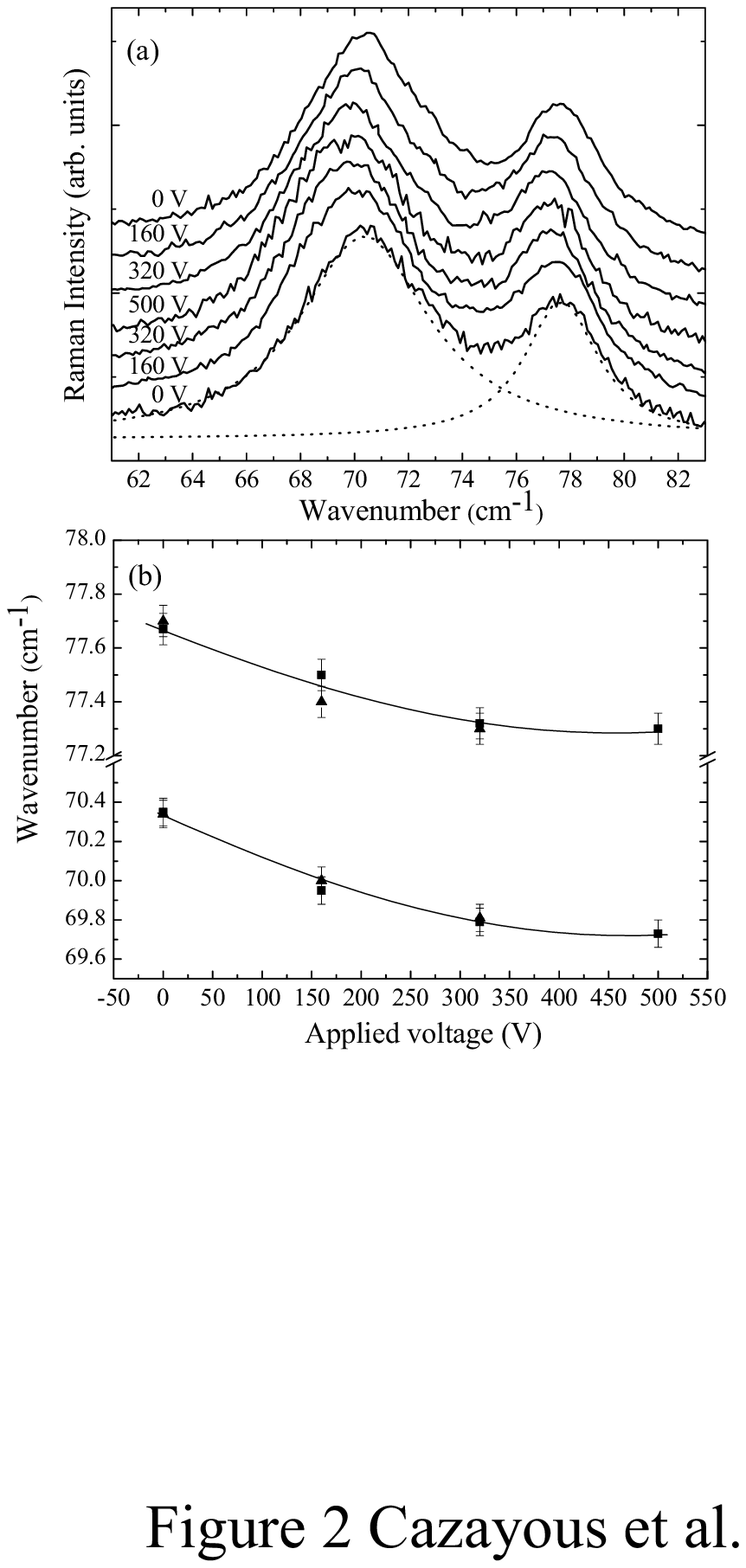}\\
\caption{\label{Figure2} (a) Raman spectra taken at different applied voltages from 0 to 500 V. 
The Lorentzian functions used for the fit of the peaks are shown as a dotted line. (b) 
Peak position versus the applied voltage. 
The lines serves as guide to the eye.}
\end{figure}

Figure~\ref{Figure2}(a) presents the Raman spectra at 300 K in the range of 
the first-order E(TO) and E(LO) modes recorded for different voltages of the external applied field. 
No frequency shift has been observed for the other vibration modes mainly due to their natural Full Width at 
Half Maximum at room temperature.
The spectra are unpolarized in order to observe the two modes. 
The applied voltage is raised up to 500 V which represents a field of 5.10$^7$V/m.
Based on a short-range force constant model, calculations have shown the contribution of the Bi-O bonds in the first-order E(TO) modes.\cite{Singh2} 
In Fig.~\ref{Figure2}(b), the observed frequency shifts of these peaks are reversible under the applied electric field and saturates for a voltage over 500V. The reversibility of the frequency shift is only observed for around 5 operation cycles before the breakdown of the sample resistance.
\par
The phonon frequency shift of the the first-order E(TO) points out the piezoelectric strain/stress caused by the applied electrical field. 
The softening of theses peaks is due to tensile stress leading to the expansion of the crystal. 
According to the theory of Raman spectroscopy\cite{Briggs, Sakaida}, the Raman shift of the E modes is related to 
the strain $\epsilon_{ij}$ via phonon deformation potentials $\alpha$, $\beta$ and $\gamma$
\begin{equation}
\Delta\omega_{E}=\alpha(\epsilon_{xx}+\epsilon_{yy})+\beta\epsilon_{zz}+
\gamma[(\epsilon_{xx}-\epsilon_{yy})^2]^{1/2}
\label{equa1}
\end{equation}
where $\epsilon_{xx}$, $\epsilon_{yy}$ and $\epsilon_{zz}$ are the normal strains on the plane of trigonal crystal and  
in the c-axis direction, respectively. For simplicity, we assume that BFO can be described by a trigonal structure and that no x-y strain is generated by the electric field applied along the z axis. The contribution to phonon frequency of the x-y strain 
has thus been neglected in Eq.~\ref{equa1}. The observed shift in the phonon frequency is then
 related to the distribution of the $E_z$ field across the crystal using the phonon deformation potential $\beta$. 
 Unfortunately, this potential is not known. Nevertheless, the electrical field along the z axis
 can be related to the strain $\epsilon_{zz}=d_{33}E_z$ in a simplified way using the linear 
 piezoelectric coefficient $d_{33}$.\cite{Felten} Taking the value for thin films reported in Ref.~\onlinecite{Wang}, the corresponding 
 strain $\epsilon_{zz}$ from the $E_z$ field is 3.10$^{-3}$ for 500~V. To get a Raman shift of 0.6~cm$^{-1}$,  the phonon deformation potential associated with the E(TO) at 70~cm$^{-1}$ is -200 cm$^{-1}$/strain unit. This value is in the range of the ones obtained for typical piezoelectrics.\cite{Briggs} 
 Using a measured or a calculated value of $\beta$, it is possible to determine $d_{33}$ in the same way. 
ne can notice that the piezoelectric response can be also determined by the phonon frequencies.\cite{Wu}
In Fig. 2(b), a reversible increase of the peak FWHM with voltage is observed. 
The broadening of the peaks indicates that the induced distorsion of the Bi-O bond is not constant along the penetration depth of the Raman probe probably due to the inhomogeneity of the applied electric field in the sample.    
\par
In summary, our measurements reveal a large TO-LO splitting in BiFeO$_3$ single 
crystal that is not observed in BiFeO$_3$ thin films. An applied electric field induces a frequency shift of the Bi-O bonds. 
This shift is related to the tensile stress due to the piezoelectric effect and allows us to estimate the 
phonon deformation potentials and the linear piezoelectric coefficient $d_{33}$.\\
\par
The authors would like to thank A. Sacuto and Y. Gallais for enlightening discussions 
and for a critical reading of the manuscript.

\newpage

\begin{table}[ht]
\parbox{8cm}{\caption{\label{assignment}
A$_1$ and E modes for BiFeO$_3$ single crystal.}}
\begin{center}
\begin{tabular}{cccccccc}
Frequency  & &  Assignment & & Frequency & &  Assignment\\ 
(cm$^{-1})$ & &   & &  (cm$^{-1})$ & &  \\
\hline
75 & & E (TO)& & 295 & & E (TO)\\ 
\hline
81 & & E  (LO) & & 348 & & E (TO)\\ 
\hline
132 & & E (TO)& & 370 & & E (TO)\\ 
\hline
145 & & A$_1$ (LO)& & 441 & &E (TO)\\ 
\hline
175.5 & & A$_1$ (LO)& & 471 & & A$_1$ (LO)\\ 
\hline
222.7 & & A$_1$ (LO)& & 523& & E (TO)\\ 
\hline
263 & & E (TO)& & 550& &A$_1$ (TO) \\ 
\hline
276 & & E (TO)& & & &  \\ 
\end{tabular}
\end{center}
\end{table}

\newpage

\newpage

\end{document}